\documentclass[final,1p,times]{elsarticle}

\usepackage{amssymb}
\usepackage{bm}
\usepackage{graphicx}
\usepackage{amsmath}
\usepackage{lipsum}
\usepackage[dvipdfmx]{color}
\usepackage{ifpdf}
\ifpdf
\usepackage[pdftex]{graphicx}
\else
\usepackage{graphicx}
\fi

\definecolor{mygray}{gray}{0.9}
\usepackage{colortbl}
\usepackage{booktabs}
\usepackage{multirow}
\usepackage{subfigure}
\begin{document}

\begin{frontmatter}



\title{Numerical modeling of photon migration in human neck based on the radiative transport equation}

\author[label1]{Hiroyuki Fujii \corref{cor1}}
\ead{fujii-hr@eng.hokudai.ac.jp}
\author[label2]{Shinpei Okawa}
\author[label3]{Ken Nadamoto}
\author[label3]{Eiji Okada}
\author[label4]{Yukio Yamada}
\author[label5]{Yoko Hoshi}
\author[label1]{Masao Watanabe}
\address[label1]{Division of Mechanical and Space Engineering, Faculty of Engineering, Hokkaido University, Kita 13 Nishi 8, Kita-ku, Sapporo, Hokkaido 060-8628, 
Japan}
\address[label2]{National Defense Medical College, 3-2 Namiki, Tokorozawa, Saitama, Japan}
\address[label3]{Department of Electronics and Electrical Engineering, Faculty of Science and Technology,
 Keio University, 3-14-1 Hiyoshi, Kohoku-ku, Yokohama, Kanagawa, Japan}
\address[label4]{Brain Science Inspired Life Support Research Center, University of Electro-Communications, 1-5-1 Chofugaoka, Chofu, Tokyo 182-8585, Japan}
\address[label5]{Preeminent Medical Photonics Education \& Reseach Center, Hamamatsu University School of Medicine, 1-20-1 Handayama, Higashi-ku, Hamamatsu, Sizuoka 431-3192, Japan}

%

\begin{abstract}
Biomedical optical imaging has a possibility of a comprehensive diagnosis of thyroid cancer in conjunction with ultrasound imaging.  
For improvement of the optical imaging, this study develops a higher order scheme for solving the time-dependent radiative transport equation (RTE) by use of the finite-difference and discrete-ordinate methods. 
The accuracy and efficiency of the developed scheme are examined by comparison with the analytical solutions of the RTE in homogeneous media. 
Then, the developed scheme is applied to describing photon migration in the human neck model. 
The numerical simulations show complex behaviors of photon migration in the human neck model due to multiple diffusive reflection near the trachea.
\end{abstract}


\end{frontmatter}

\section{Introduction}
\label{Introduction}

Thyroid cancer is one of common cancers, whose incidence increases in accordance with developments of biomedical imaging modalities \cite{Wartofsky2010}. 
There are several kinds of thyroid cancers, and most of them can be accurately diagnosed by using the ultrasound tomography and fine needle aspiration cytology to examine the cellular structures \cite{Levi2013}. 
Nevertheless, diagnosis of follicular thyroid carcinoma is still difficult, 
because neither the ultrasound tomography nor the needle aspiration cytology can differentiate between them. 
Hence, other imaging modality is essential for comprehensive diagnosis of thyroid cancer. 

Diffuse optical tomography (DOT) has a potential to accurately and non-invasively diagnose whether the tumor is benign or malignant based on the difference in optical properties \cite{Yamada2014, Gibson2005}. 
The optical properties (e.g. the absorption and scattering coefficients, and anisotropic factor) characterize 
the strength of absorption and scattering of light by a turbid medium such as biological tissue. 
Unlike a conventional image reconstruction algorithm for X-ray computed tomography, of which fundamentals are measuring line integrals of the attenuation coefficients and solving a set of integral equations for the attenuation coefficients, 
DOT needs an inversion process to reconstruct a distribution of tomographic image of the optical  properties inside the tissues based on a mathematical model describing the diffusive nature of  photon migration in scattering media \cite{Arridge1999}. 
Hence, for obtaining a high quality image of DOT, an accurate photon migration model is essential, and an efficient scheme for solving the model is required. 

Mainly, two types of the governing equations for photon migration have been used: 
the radiative transport equation (RTE) and the diffusion equation (DE). 
The RTE can accurately describe photon migration in wide ranges of time-length scales, 
and numerical schemes for solving the RTE have extensively been developed by using the finite-difference \cite{Klose2002, Klose2005}, finite-element \cite{Abdoulaev2003}, and finite-volume methods \cite{Marin2014}. 

However, their computational loads for solving the RTE are heavy because the RTE is an integro-differential equation with many independent variables. 
Due to this reason, the RTE has usually been applied to small-sized media such as human fingers  and rats. 
Meanwhile, the DE can be applied to large-sized media (e.g. the human brain) \cite{Schweiger1993, Gao2002, Okawa2011} because the approximation can decrease the number of the independent  variables to reduce the computational loads. 
However, it is well understood the DE is valid in spatial and temporal regions after photons undergo a sufficient number of scattering events. 
In non-scattering or void regions, the DE is invalid and fails to describe photon migration \cite{Yoo1990, Hielsher1998, Yuan2009}. 

The human neck is a large-sized and inhomogeneous medium consisting of the trachea, 
arteries, veins, muscles, bones, adipose tissue, and so on. 
The trachea is a void region where photons travel straight without being absorbed and scattered. 
Also, the arteries and veins are highly absorbing regions, where photons are totally absorbed before photons undergo a sufficient number of scattering events for the DE to be valid. 
Considering the above-mentioned characteristics of the human neck, 
an efficient scheme for solving the RTE is still in need of further development. 

This paper develops the numerical scheme for solving the RTE using the finite-difference and discrete-ordinate methods 
(the 3rd order upwind and the 4th order Runge-Kutta methods). 
At first, the validity of the developed scheme is confirmed for homogeneous media by comparison with  analytical solutions. 
After the validation, the developed scheme is applied to investigation of photon migration in the human neck. 

The following section provides an explanation of the numerical method based on the RTE. 
Section 3 provides the numerical results for a 2D homogeneous square medium and an inhomogeneous medium modeled from an MR image of the human neck. 
Finally, conclusions are given in section 4. 
\section{Light propagation model}

\section{Numerical method}
\subsection{Radiative transport equation and numerical scheme}

2D photon migration in turbid media such as biological tissues is accurately formulated by the RTE \cite{Chandra1960}, 
\begin{equation}
	\left[ \frac{\partial}{v\partial t}+\bm{\Omega}\cdot \nabla+\mu_a(\bm{r})+\mu_s(\bm{r})\right]
	I(\bm{r}, \bm{\Omega}, t)
	= \mu_s(\bm{r}) \int_{2\pi} d\bm{\Omega}' P(\bm{r}, \bm{\Omega} \cdot \bm{\Omega}')I(\bm{r}, \bm{\Omega}', t) 	+ q(\bm{r}, \bm{\Omega}, t),
	\label{eq:2_1_1}
\end{equation}
where $I(\bm{r}, \bm{\Omega}, t)$ [W/cm rad] represents the intensity, which describes the photon energy flow 
as a function of position, $\bm{r}=(x, y)$ [cm], angular direction, 
$\bm{\Omega}=(\Omega_x, \Omega_y)=(\cos \theta, \sin \theta)$ with angle, $\theta$ [rad], and time, $t$ [ps]. 
$\mu_a(\bm{r})$ [1/cm] and $\mu_s(\bm{r})$ [1/cm] are the absorption and scattering coefficients, respectively, 
$v$ [cm/ps] is the velocity of light in the turbid medium, $P(\bm{r}, \bm{\Omega} \cdot \bm{\Omega}')$ [1/rad] is the scattering phase function with $\bm{\Omega}$ and $\bm{\Omega}'$ denoting the scattered and incident directions, respectively, and $q(\bm{r}, \bm{\Omega}, t)\; \rm{[W/cm^2 \;rad]}$ is a source. 
In this study, $P(\bm{r}, \bm{\Omega} \cdot \bm{\Omega}')$ is given by the Henyey-Greenstein function \cite{Henyey1941} in two dimensions \cite{Heino2003}, 
\begin{equation}
	P(\bm{r}, \bm{\Omega} \cdot \bm{\Omega}')= \frac{1}{2 \pi} \frac{1-\{g(\bm{r})\}^2}{1+
		\{ g(\bm{r}) \}^2-2g(\bm{r}) \bm{\Omega} \cdot \bm{\Omega}'}, 
	\label{eq:2_1_2}
\end{equation}
where $g(\bm{r})$ represents the anisotropic factor defined as the average cosine of $P(\bm{r}, \bm{\Omega} \cdot \bm{\Omega}')$. 
For simplicity, $q(\bm{r}, \bm{\Omega}, t)$ is given by an isotropic delta function. 
The boundary condition under the refractive-index mismatching is employed with the reflectivity, $R(n,  \bm{\Omega} \cdot \hat{\bm{e}}_n(\bm{r}_b))$, calculated by the Fresnel's law and Snell's law, where $n$ denotes the refractive index of the medium and $\hat{\bm{e}}_n(\bm{r}_b)$ denotes the outward normal vector at the boundary position, $\bm{r}_b$ \cite{Klose2005}. 

In this study, the RTE is numerically solved based on the finite-difference and discrete-ordinate methods. 
For numerical discretization, $x$, $y$, $\theta$, and $t$ are divided into $x_i=i \Delta x$ $(i \in \{0, \;\cdots, \;N_x\})$, $y_j=j \Delta y$ $(j \in \{0, \;\cdots, \;N_y\})$, 
$\theta_k=k \Delta \theta$ $(k \in \{0, \;\cdots, \;N_{\theta} \})$, and $t_m=m \Delta t$ 
$(m \in \{0, \;\cdots, \;N_t\})$ with constant step sizes of 
$\Delta x$, $\Delta y$, $\Delta \theta$, and $\Delta t$, respectively, 
and numbers of grid nodes and timesteps, $N_x$, $N_y$, $N_{\theta}$, and $N_t$, respectively. 
In the same manner, $\bm{\Omega}$ is discretized as $(\Omega_{kx}, \Omega_{ky})=(\cos \theta_k, \sin \theta_k)$, 
and $I(\bm{r}, \bm{\Omega}, t)$ at the grid node and timestep  is expressed as $I_{ijk}^m$. 

For the finite-difference discrete-ordinate methods, previous papers mostly employ the 1st order scheme: 
the 1st order upwind (1UW) scheme for $\bm{r}$, the extended trapezoidal rule for $\bm{\Omega}$, 
and the forward Euler method (FE) for $t$ \cite{Fujii2014JQSRT, Klose1999}. 
The 1st order scheme is stable and accurate if the spatial and temporal step sizes, $\Delta x$ and $\Delta t$, are sufficiently fine, but the fine step sizes result in heavy computational loads. 
To reduce the computational loads, this study employs the 3rd order upwind (3UW) scheme and 4th order Runge-Kutta (4RK) method. 

In the 3UW, the advection term in the $x$-direction, $\Omega_x \partial I (\bm{r}, \bm{\Omega})/\partial x$, is discretized as 
\begin{equation}
	\Omega_x \partial I (\bm{r}, \bm{\Omega})/\partial x \sim
\begin{cases}
\frac{\Omega_{kx}}{6 \Delta x} \left[ 2 I_{i+1 j k}+3 I_{ijk}-6I_{i-1jk}+I_{i-2jk} \right]& \Omega_{kx} \geq 0 \\
\frac{\Omega_{kx}}{6 \Delta x} \left[ - I_{i+2 j k}+6 I_{i+1jk}-3I_{ijk}-2I_{i-1jk} \right] & \Omega_{kx} < 0
\end{cases}, 
	\label{eq:2_1_3}
\end{equation}
for the internal nodes. 
For the grids near the boundary, the 1UW scheme is employed. 
The advection term in the $y$-direction, $\Omega_y \partial I (\bm{r}, \bm{\Omega})/\partial y$, is  also discretized in the same manner as in the $x$-direction. 
The 3UW is slightly unstable compared to the 1st order scheme, 
and oscillations appear at early arriving times and at positions very  close to the source. 
In this study, the media is so large that the oscillations are negligibly small. 

Based on the extended trapezoidal rule, the scattering integral is calculated as 
\begin{equation}
	\int_{2\pi} d\bm{\Omega}' P(\bm{r}, \bm{\Omega}\cdot\bm{\Omega}')I(\bm{r}, \bm{\Omega}') \sim 
		\sum_{k'=1}^{N_{\theta}} w_{k'}f_{k'}P_{kk'} I_{ijk'}, 
	\label{eq:2_1_4}
\end{equation}
where $w_{k'}=2\pi/N_{\theta}$ is a weight and $f_{k'}$ is a renormalizing factor. 
To satisfy the normalization of the scattering phase function $\int_{2\pi} d\bm{\Omega}' P=1$, 
$f_{k'}$ is formulated with a slight modification of the Liu's renormalization \cite{Liu2002}, 
$1/f_{k'}=\sum_{k'=1}^{N_{\theta}} w_{k'} P_{kk'}$. 

In the matrix notation, the discretized RTE except for the time variable, $t$, is given as 
an algebraic equation, 
\begin{equation}
	\frac{\partial}{v\partial t}\bm{I} + \bm{A}\bm{I}=\bm{P}\bm{I}+\bm{Q}, 
	\label{eq:2_1_5}
\end{equation}
where a matrix $\bm{A}$ represents the second, third, and fourth terms in the left-hand side of Eq. (\ref{eq:2_1_1}), 
a matrix $\bm{P}$ represents the scattering integral (Eq. (\ref{eq:2_1_4})) multiplied by $\mu_s(\bm{r})$, 
vectors $\bm{I}$ and $\bm{Q}$ are the intensity and source, respectively. 

Based on the 4RK, the temporal change in the intensity is calculated as 
\begin{equation}
	\bm{I}^{m+1} \sim \bm{I}^{m}+\frac{1}{6}(\bm{k}_1+2\bm{k}_2+2\bm{k}_3+\bm{k}_4), 
	\label{eq:2_1_6}
\end{equation}
where $\bm{k}_l$($l=1$, 2, 3, 4) are given by the following equations, 
\begin{eqnarray}
	\bm{k}_1 &=&v \Delta t (-\bm{A}+\bm{P})\bm{I}^{m}, \nonumber \\ 
	\bm{k}_2 &=&v \Delta t (-\bm{A}+\bm{P})(\bm{I}^{m}+0.5\bm{k}_1), \nonumber \\
	\bm{k}_3 &=&v \Delta t (-\bm{A}+\bm{P})(\bm{I}^{m}+0.5\bm{k}_2), \nonumber \\
	\bm{k}_4 &=&v \Delta t (-\bm{A}+\bm{P})(\bm{I}^{m}+\bm{k}_3).
	\label{eq:2_1_7}
\end{eqnarray}
Because the explicit method is used for 4RK, 
it should satisfy the Courant-Friedrichs-Lewy (CFL) condition, $\Delta t \le \Delta x/2v$, for stable solutions \cite{Klose1999, Guo2001}. 

In this study, parallel programming is implemented with a 24 threads computer 
(Intel Xeon X5690 @3.47GHz) by using OpenMP, 
which is a portable and shared-memory programming scheme. 

\subsection{Geometry, optical properties, and step sizes in discretization}
\begin{figure}[tb]
\centering
(a)
\hspace{-0.9cm}
\mbox{\raisebox{0.2cm}{\includegraphics[scale=0.24]{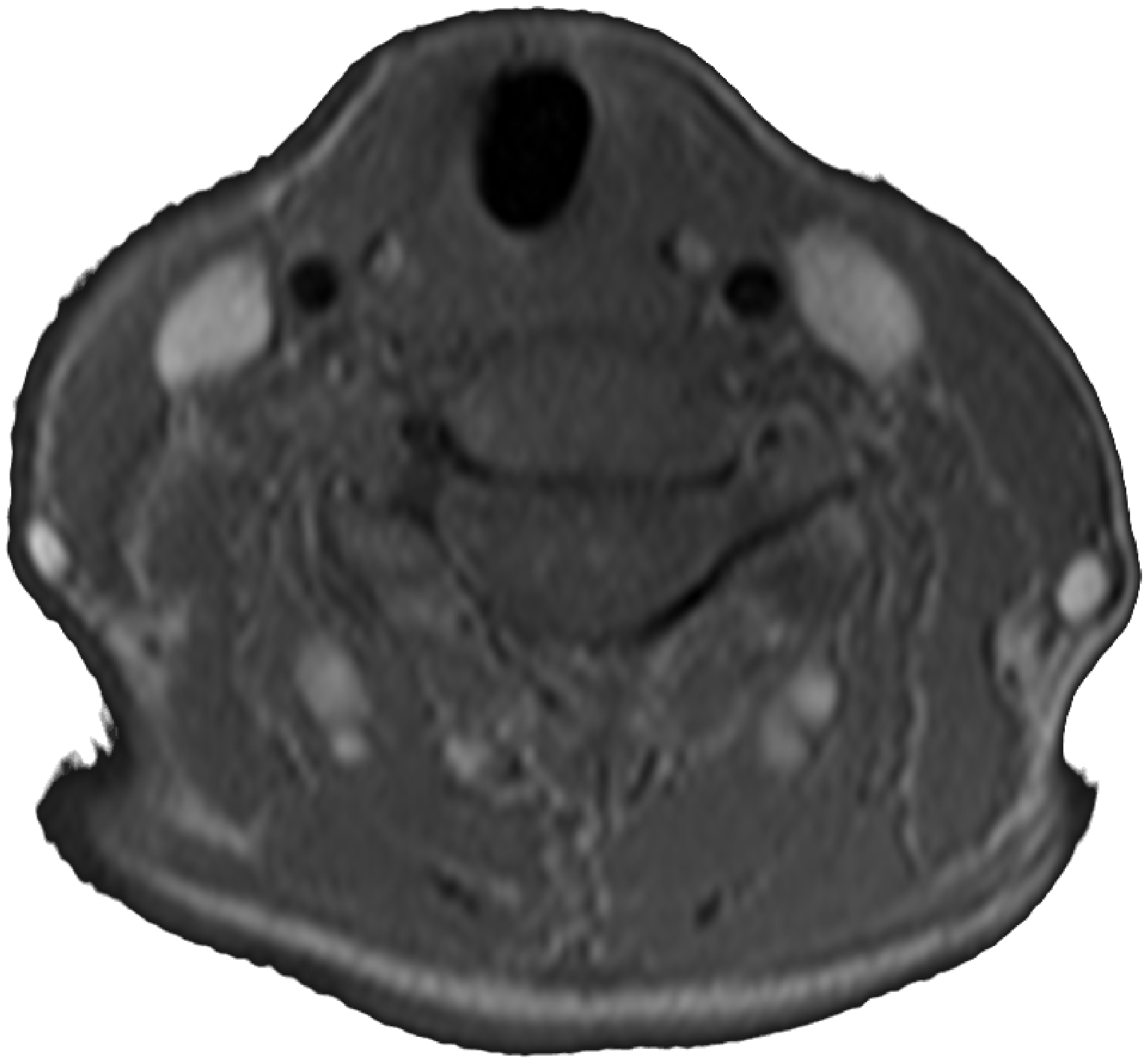}}}
\hspace{0.2cm}
(b)
\mbox{\raisebox{0.0cm}{\includegraphics[scale=0.35]{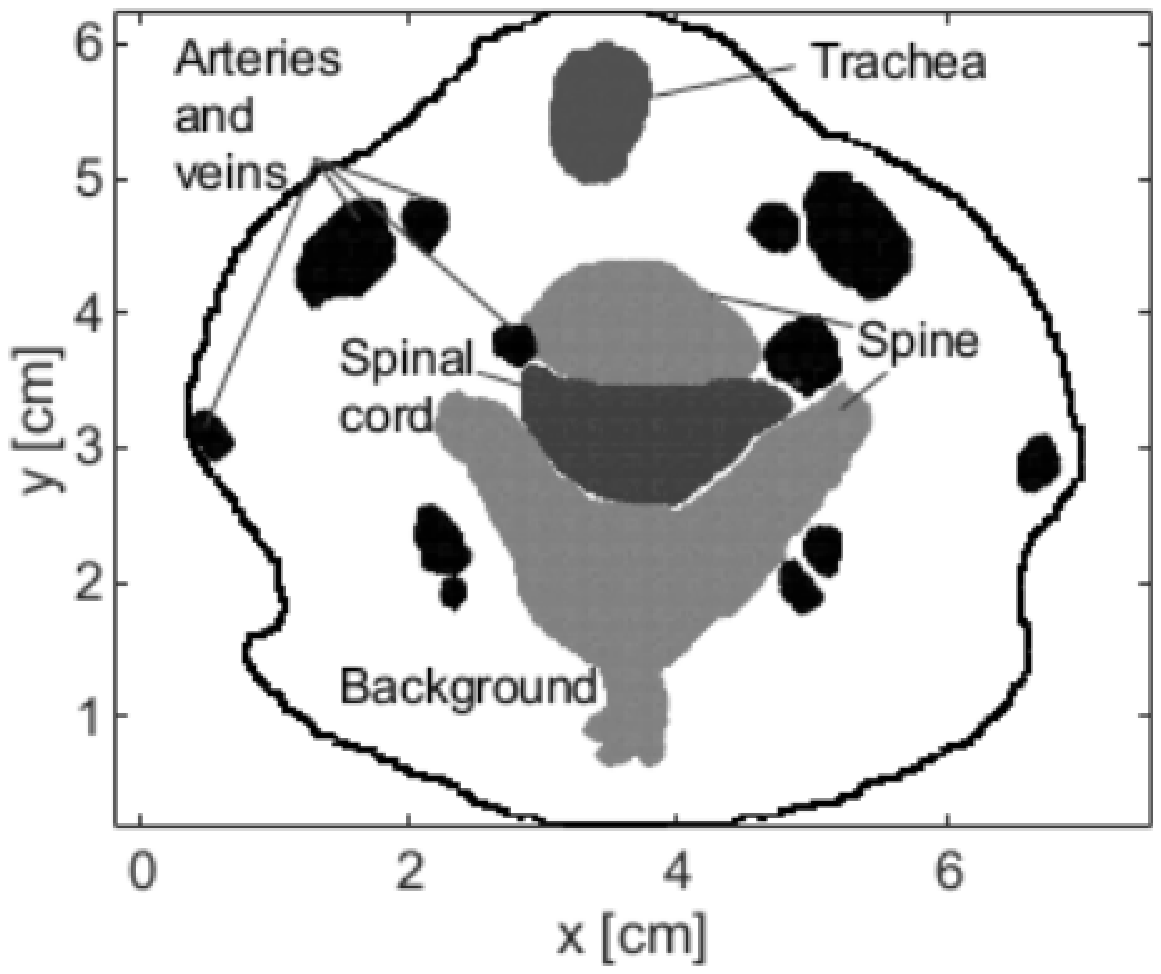}}}
\hspace{-1.0cm}
\caption{(a) MR image of the human neck, (b) primary organs after the segmentation}
\label{fig1}
\end{figure}

To construct the human neck model, the study uses an MR image of the human neck as shown in Fig. \ref{fig1} (a). 
Segmentation of the MR image is performed, and primary organs in the human neck such as common carotid artery, internal jugular vein, trachea, spinal cord, and spine are extracted as shown in Fig. \ref{fig1} (b). 
Other organs such as human muscle, bone, and adipose tissue are collected to form a homogeneous background. 
Then, irregular boundaries of the organs and background in the image are mapped with regular grids for numerical calculations. 
At the surface of the human neck, the Fresnel's law is employed for the boundary condition which describes the reflection and refraction at the boundary due to the difference in the refractive index between the medium and surrounding air. 
Meanwhile, we assume that the refractive index inside the medium is homogeneous. 
So neither reflection nor refraction takes places at the interfaces between different organs, and no boundary condition is used at the interfaces. 
The optical properties of the organs and background tissues in the human neck in the near-infrared wavelength range are listed in Table \ref{tab:1}, referring to \cite{Bashkatov2011, Dehaes2011}, 
and the properties of the thyroid gland is considered to be the same as those of the background. 
Strictly speaking, the arteries have different values of the properties from the veins, but 
it is simply assumed that the both have the same optical properties. 
Except the trachea, the tissues are assumed to have a constant value of $g=0.9$, corresponding to  highly forward-peaked scattering. 

Appropriate values of the step sizes are dependent on the optical properties; 
$\Delta x$ and $\Delta t$ are inversely proportional to $\mu'_s=\mu_s(1-g)$ and $\mu_a$, and 
$\Delta \theta$ should be smaller when $g$ approaches to one. 
In each case, we set the appropriate values of the step sizes. 
\begin{table}
\centering 
\caption{Optical properties of the primary organs and background in the human neck \cite{Bashkatov2011, Dehaes2011}}
\label{tab:1}
\begin{tabular}{lccc}
\hline
  & $\mu_a$[1/cm]
  & $\mu_s$[1/cm]
  & $g$\\
\hline
Artery and vein & 4.76 & 675 & 0.9 \\
Trachea & 0.0001 & 0.0001 & 0.0 \\
Spinal cord & 0.17 & 882 & 0.9 \\
Spine & 0.25 & 148 & 0.9 \\
Background & 0.30 & 80.0 & 0.9 \\
\hline
\end{tabular}
\end{table}
\section{Numerical results}
\subsection{Validation of the developed numerical scheme}

Before a discussion on the results of photon migration in the human neck model, 
the developed numerical scheme is validated for the case of a homogeneous 2D square medium with  a side of 4.0 cm by comparison with analytical solutions for infinite media. 
When the source is located inside the medium far from the boundary, 
the numerical solutions of the intensities at the positions far from the boundary are less influenced by the boundary. 
Then, the comparisons of the numerical solutions for finite media with the analytical solutions for infinite media are reasonable. 
In this subsection, an isotropic impulse point source is incident at ($x$, $y$) = (1.5 cm, 2.0 cm), and the intensities at (2.5 cm, 2.0 cm) are calculated. 

At first, the case of isotropic scattering ($g=0.0$) is examined because 
the exact analytical solution of the RTE has been given by Paasschens \cite{Paasschens1997} as follows, 
\begin{equation}
	\Phi_{ER}(\rho, t) 
	=v\frac{\delta(vt-\rho)}{2\pi \rho}e^{-\mu_t v t}
	+\frac{v \mu_s}{2\pi} \frac{\exp(\mu_s \sqrt{(vt)^2-\rho^2})}{\sqrt{(vt)^2-\rho^2}}e^{-\mu_t v t}\Theta(vt-\rho), 
	\label{eq:3_1}
\end{equation}
where $\Phi$ is the fluence rate defined as $\int_{2\pi} d\bm{\Omega} I$, $\Theta$ is the Heaviside step function, $\mu_t=\mu_a+\mu_s$, and $\rho$ is the distance from the source. 
Also, the analytical solution of the DE \cite{Chandra1943}, $\Phi_{DE}$, is given as, 
\begin{equation}
	\Phi_{DE}(\rho, t) = \frac{1}{4\pi D t} e^{-\mu_a v t} e^{-\frac{\rho^2}{4Dvt}}, 
	\label{eq:3_2}
\end{equation}
where the diffusion coefficient $D$ is given as $[2(1-g)\mu_s]^{-1}$ for the 2D time-dependent case \cite{Furutsu1994, Pierrat2006}. 
The absorption and scattering coefficients are given as the same as those of either 
(a) the background: $\mu_a=0.30$ $\rm cm^{-1}$ and $\mu_s=80.0$ $\rm cm^{-1}$, 
or (b) the artery: $\mu_a=4.76$ $\rm cm^{-1}$ and $\mu'_s=67.5$ $\rm cm^{-1}$ as listed in Table 1. 
The former and latter cases correspond to weakly and strongly absorbing media, respectively.  
The step sizes are given as (a) $\Delta x=\Delta y=0.02$ cm, $\Delta \theta=0.13$ rad, and $\Delta t=0.3$ ps, and (b) $\Delta x=\Delta y=0.005$ cm, $\Delta \theta=0.13$ rad, and $\Delta t=0.07$ ps, respectively. 

\begin{figure}[tb]
\centering
(a)
\includegraphics[scale=0.365]{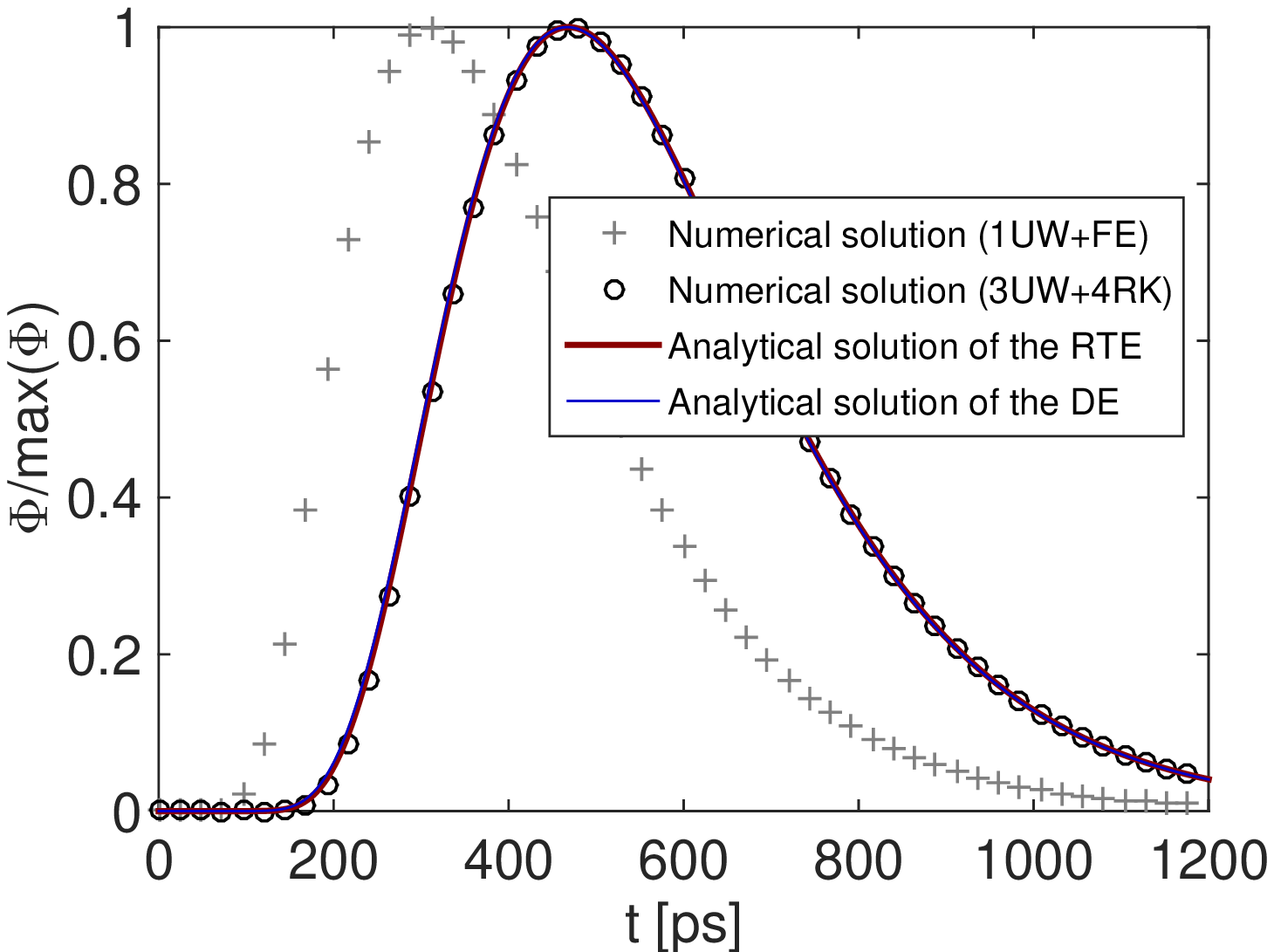}
\hspace{1.0 cm}
(b)
\includegraphics[scale=0.36]{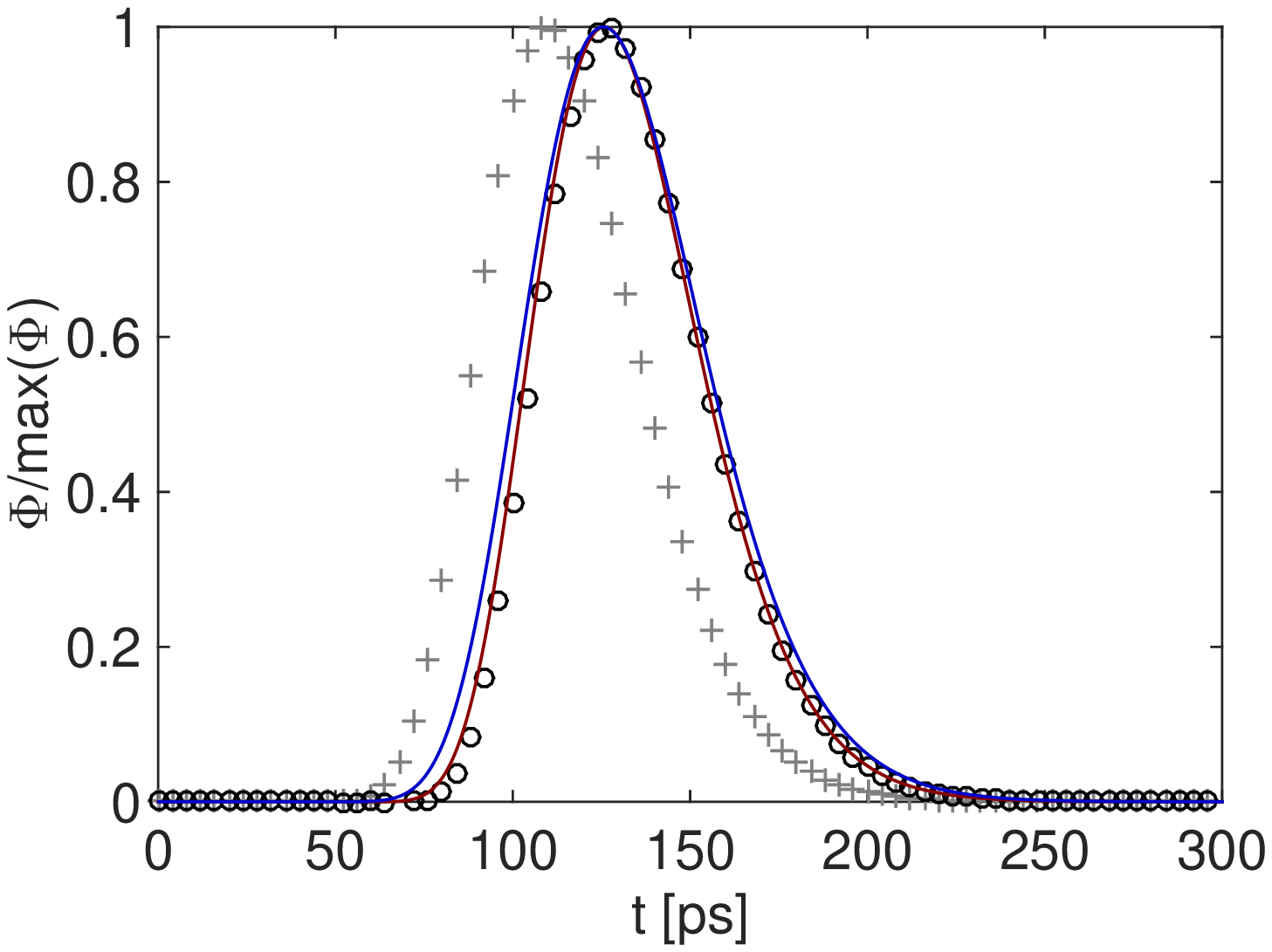}
\caption{Normalized time-resolved profiles of $\Phi$ for the RTE and the DE in the case of the isotropic scattering ($g=0.0$) at $\rho=1.0$ cm: pluses and circles represent the numerical solution based on the 1UW+FE, and the 3UW+4RK, respectively, and red and blue solid curves the analytical solutions, $\Phi_{ER}$ and $\Phi_{DE}$, respectively. 
The optical properties are given as (a) $\mu_a=0.30\;\rm cm^{-1}$, $\mu_s=80.0\;\rm cm^{-1}$, and $g=0.0$ and (b) $\mu_a=4.76\;\rm cm^{-1}$, $\mu_s=67.5\;\rm cm^{-1}$, and $g=0.0$. } 
\label{fig2}
\end{figure}

Figure \ref{fig2} shows the time-resolved profiles of $\Phi$ calculated numerically and analytically at $\rho=1.0$ cm. 
As shown in Fig. \ref{fig2} (a), the numerical solution of the RTE based on the 3UW+4RK agrees very well with the analytical solutions in the case of the weakly absorbing medium. 
Meanwhile, the numerical solution based on the 1UW+FE deviates from the analytical solutions. 
This result indicates that in the case of the 1st order scheme, the current values of $\Delta x$ and $\Delta t$ are inappropriate, and smaller values of them are necessary to obtain good agreement with  the analytical solutions. 
Note that $\Phi_{DE}$ is in good agreement with $\Phi_{ER}$ and the numerical solution based on the higher order scheme. 
This is because $\rho$ is sufficiently long for the DE to hold. 

The results of the case of the strongly absorbing medium are shown in Fig. \ref{fig2} (b), where $\Phi$ has sharp peaks at earlier times than in Fig. \ref{fig2} (a) due to stronger absorption. 
Similarly to the case of the weakly absorbing medium (Fig. \ref{fig2} (a)), 
the numerical solution based on the 3UW+4RK agrees very well with $\Phi_{ER}$, 
meanwhile the solution based on the 1UW+FE deviates from $\Phi_{ER}$. 
It is seen that the curve of $\Phi_{DE}$ has a similar shape to that of $\Phi_{ER}$ even for the strongly absorbing medium, 
although the peak time and width of the curve of $\Phi_{DE}$ are slightly different from those of $\Phi_{ER}$. 

\begin{figure}[tb]
\centering
(a)
\includegraphics[scale=0.32]{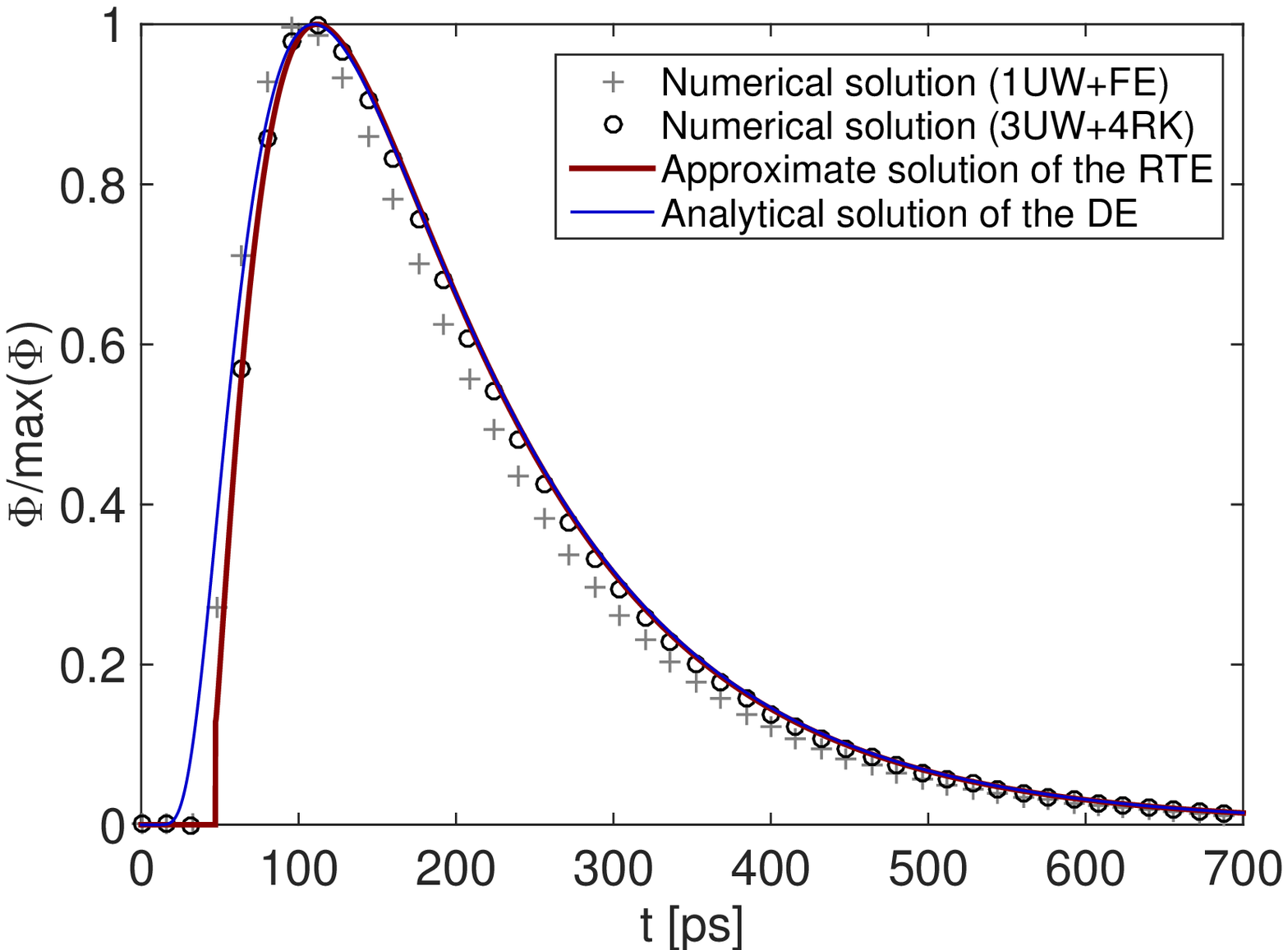}
\hspace{1.0 cm}
(b)
\includegraphics[scale=0.36]{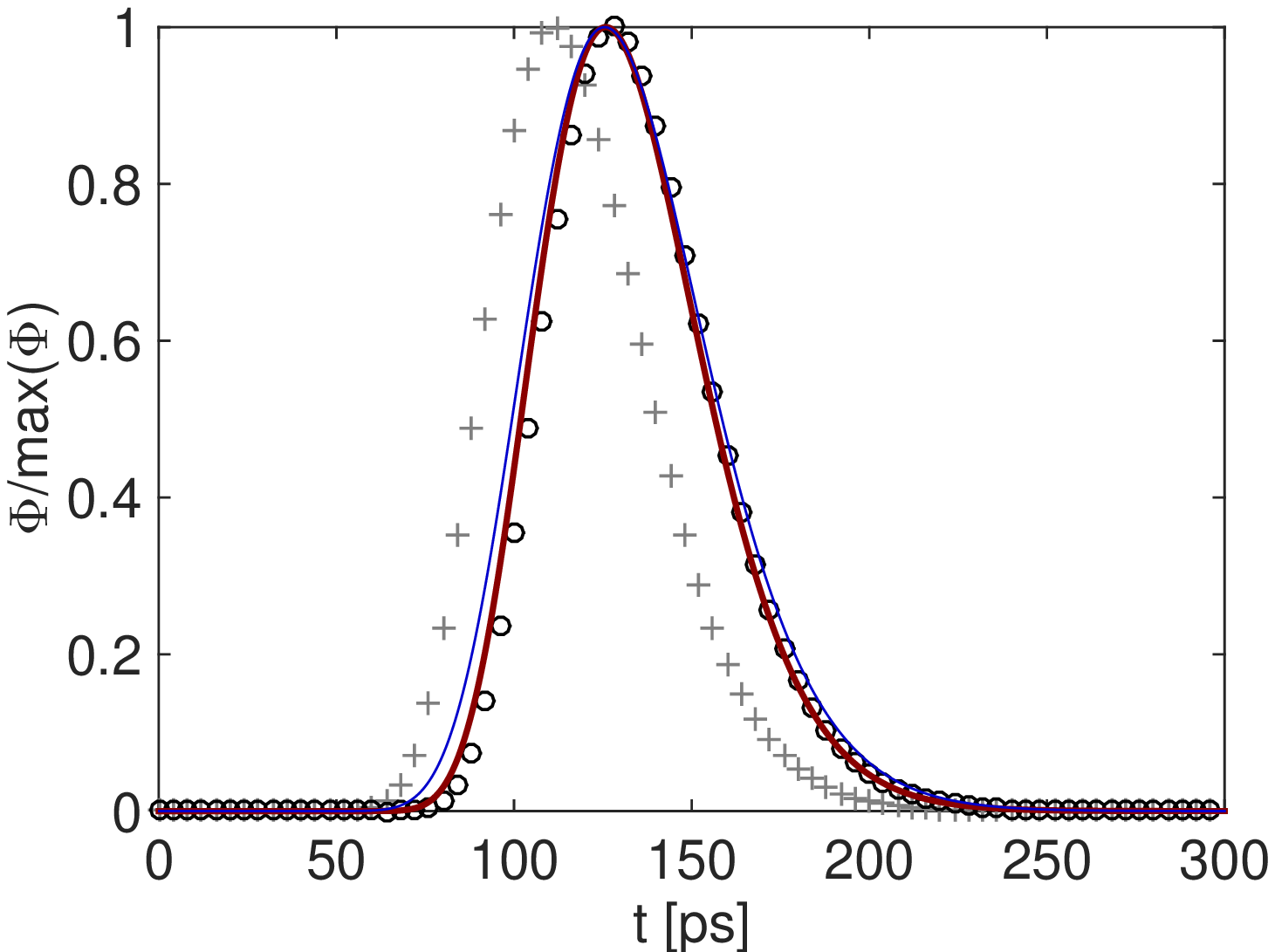}
\caption{Anisotropic scattering case ($g=0.9$) of $\Phi$ at $\rho=1.0$ cm: a red solid curve represents the approximate solution of the RTE, $\Phi_{AR}$. (a) $\mu_a=0.30\;\rm cm^{-1}$, $\mu_s=80.0\;\rm cm^{-1}$, and  $g=0.9$ and (b) $\mu_a=4.76\;\rm cm^{-1}$, $\mu_s=675\;\rm cm^{-1}$, and $g=0.9$. Other details are the same as in Fig. \ref{fig2}. }
\label{fig3}
\end{figure}

In the case of anisotropic scattering media ($g\neq0$), a derivation of exact analytical solutions of the RTE is  still a difficult problem 
even in infinite homogeneous media. 
In this study, an approximate RTE solution, $\Phi_{AR}$, proposed by Martelli \cite{Martelli2007} is used to be compared with the numerical solutions.  
The basic assumption for $\Phi_{AR}$ is that $\mu_s$ and $\mu_t$ in Eq. (\ref{eq:3_1}) can be replaced by 
$\mu'_s=\mu_s(1-g)$ and $\mu'_t=\mu'_s+\mu_a$, respectively. 
Again, two sets of the optical properties of the medium are considered: (a) $\mu_a=0.30$ $\rm cm^{-1}$, $\mu_s=80.0$ $\rm cm^{-1}$, and $g=0.9$ of the background, 
and (b) $\mu_a=4.76$ $\rm cm^{-1}$, $\mu_s=675$ $\rm cm^{-1}$, and $g=0.9$ of the artery. 
Other conditions are the same as those in the case with the isotropic scattering media. 
As shown in Fig. \ref{fig3} (a) in the case of the weak absorption, the numerical solution of the RTE based on the 3UW+4RK agree well with $\Phi_{AR}$. 
However, in details, a small difference between them is seen during the period from about 100 ps to 400 ps. 
Even if the step sizes are finer, the numerical solution is found to be unchanged. 
Thus, this small difference could be the error in $\Phi_{AR}$ due to the approximation. 
In Fig. \ref{fig3} (a), the deviations of the numerical solution based on the 1UW+FE are small  compared with that in the case of isotropic scattering. 
Because the appropriate value of $\Delta x$ is inversely proportional to $\mu'_s$, 
it is much larger in the anisotropic scattering case with $\mu'_s=8.0\; {\rm cm^{-1}}$ than in the isotropic scattering case with $\mu'_s=80\; {\rm cm^{-1}}$. 
This is the reason why the 1UW+FE performs better in the anisotropic scattering case. 
 
Figure \ref{fig3} (a) shows that $\Phi_{DE}$ is in agreement with the approximate RTE solution at the  peak time ($t \sim$100 ps) and the decay period ($t>100$ ps). 
At early arriving time ($t<100$ ps), meanwhile, $\Phi_{DE}$ deviates from the approximate RTE solution. 
This deviation comes from the difference in the speed of light between the RTE and the DE. 
In the RTE, the speed of light is given as $v$, while in the DE it is infinite \cite{Guo2001}. 

The strongly absorbing and anisotropic scattering case is shown in Fig. \ref{fig3} (b). 
The curves of $\Phi$ in the case of anisotropic scattering are the same as those in the case of isotropic scattering (Fig. \ref{fig2} (b)). 
This is maybe because the curves are mostly determined by $\mu'_s$, and the values of $\mu'_s$ are the same for the two cases. 

From the results shown in Figs. \ref{fig2} and \ref{fig3}, 
the developed schemes is validated. 

\subsection{Photon migration in the human neck model}
\begin{figure}[tb]
\centering
\subfigure[46.2 ps]{
\includegraphics[scale=0.26]{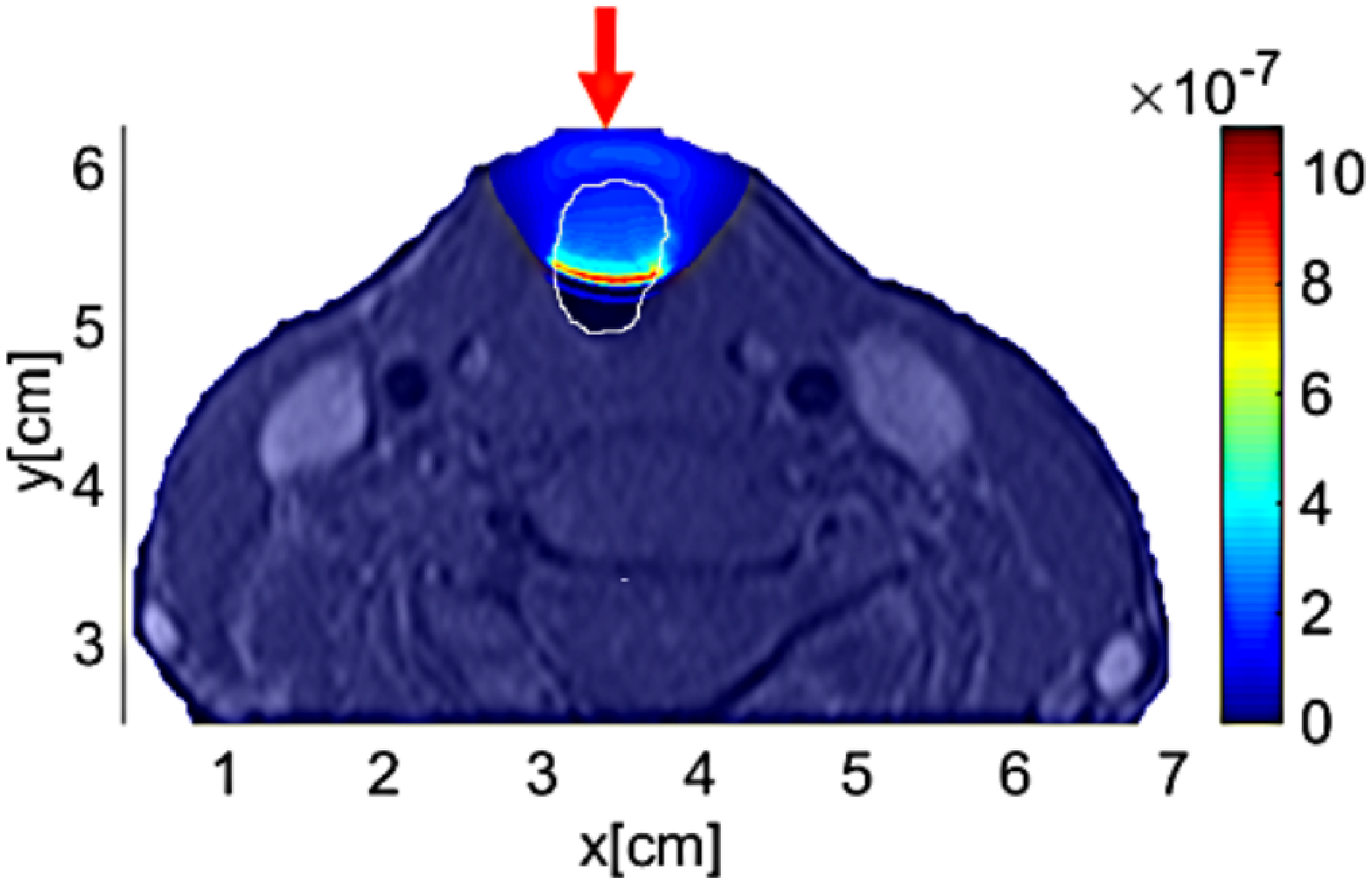}
}
\subfigure[79.9 ps]{
\includegraphics[scale=0.26]{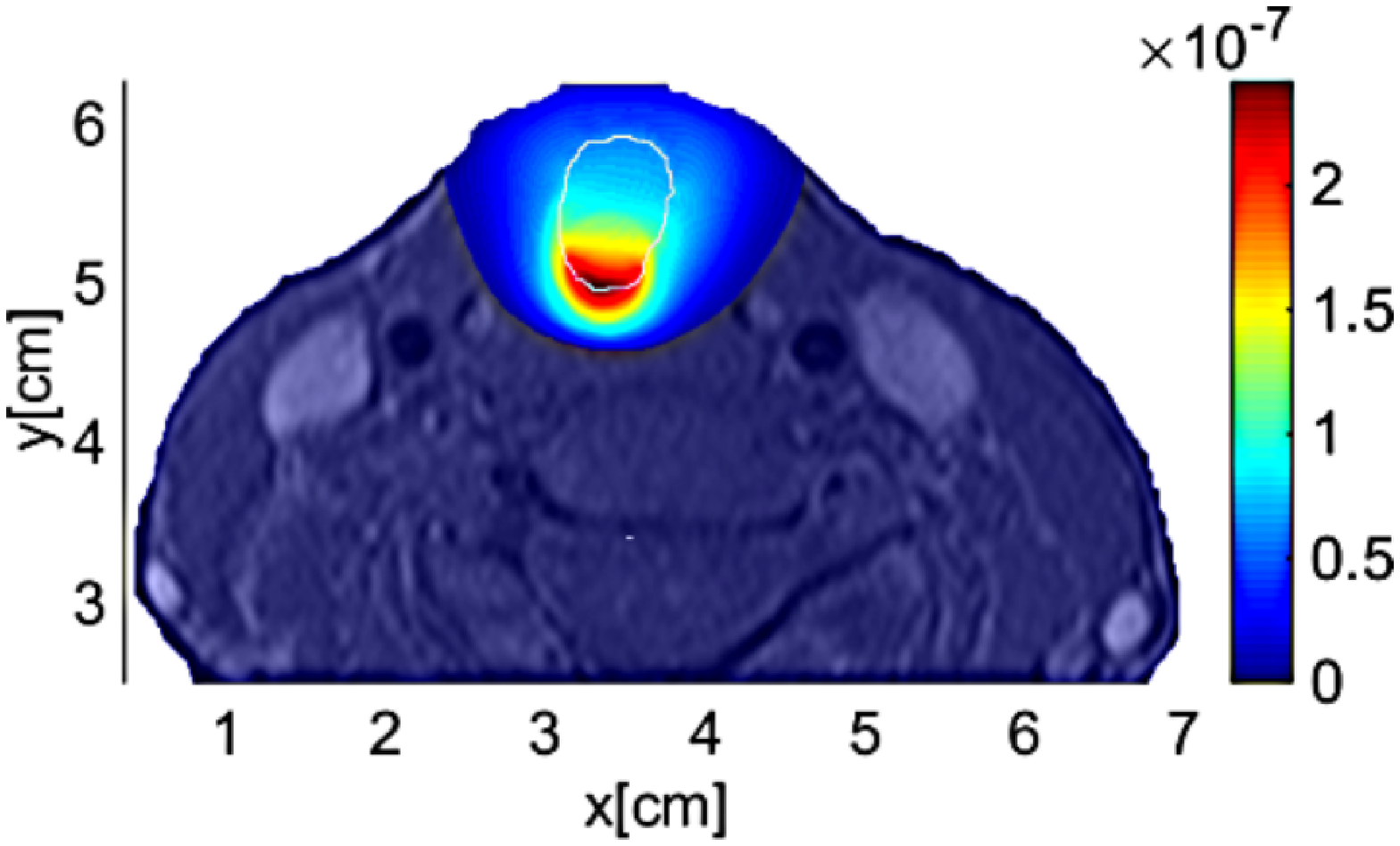}
}
\subfigure[163.9 ps]{
\includegraphics[scale=0.26]{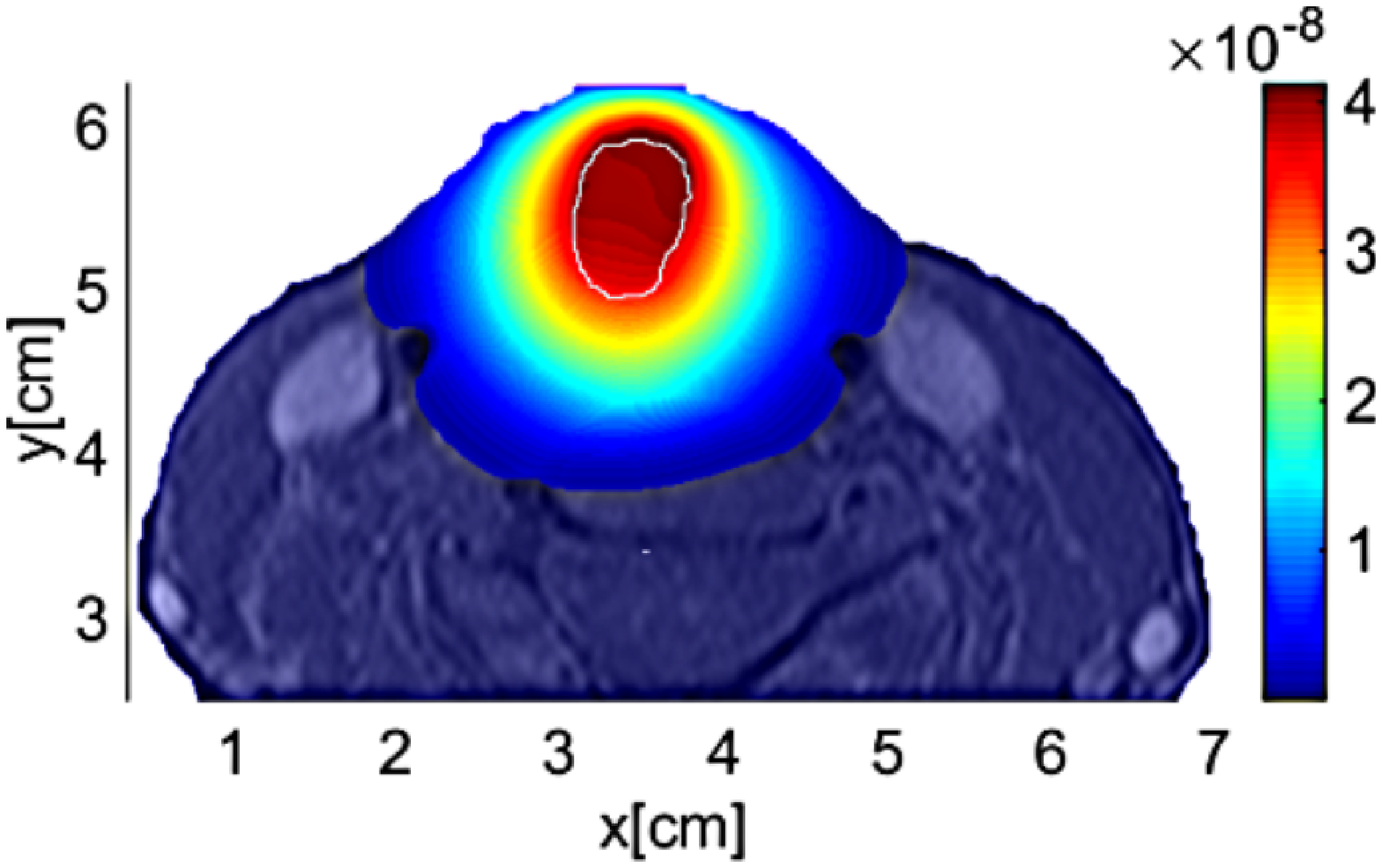}
}\\
\subfigure[285.8 ps]{
\includegraphics[scale=0.26]{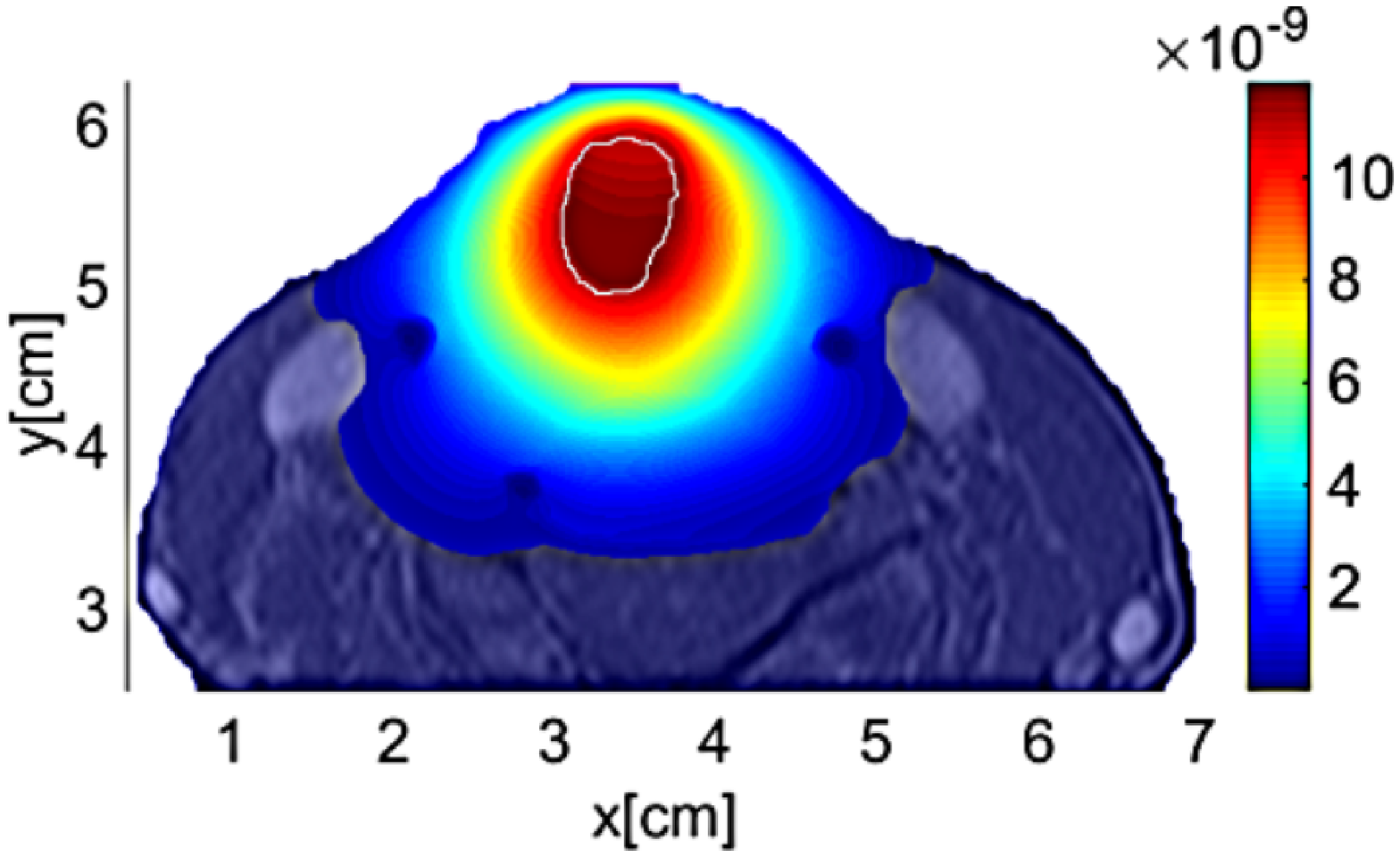}
}
\subfigure[512.8 ps]{
\includegraphics[scale=0.26]{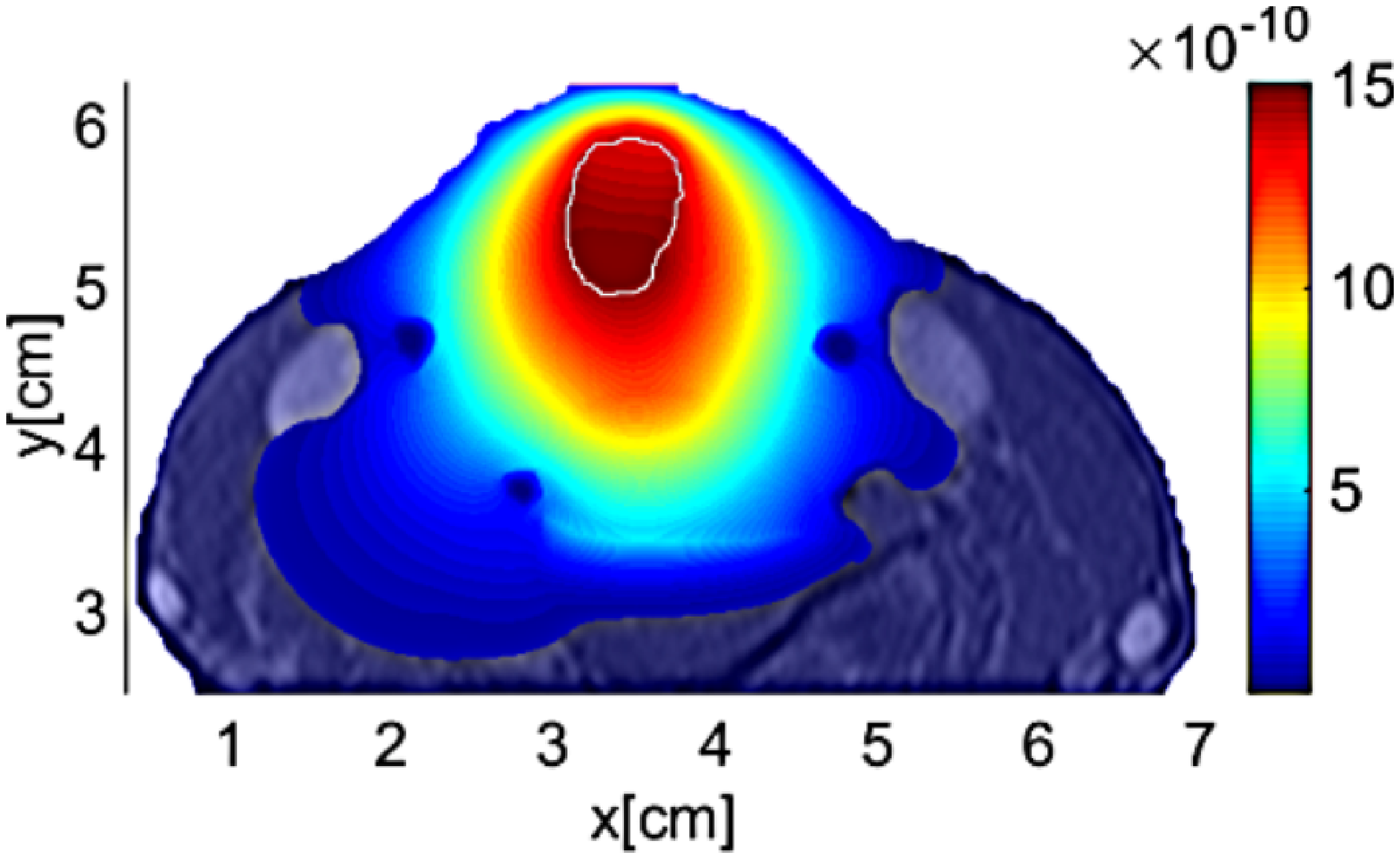}
}
\subfigure[580.0 ps]{
\includegraphics[scale=0.26]{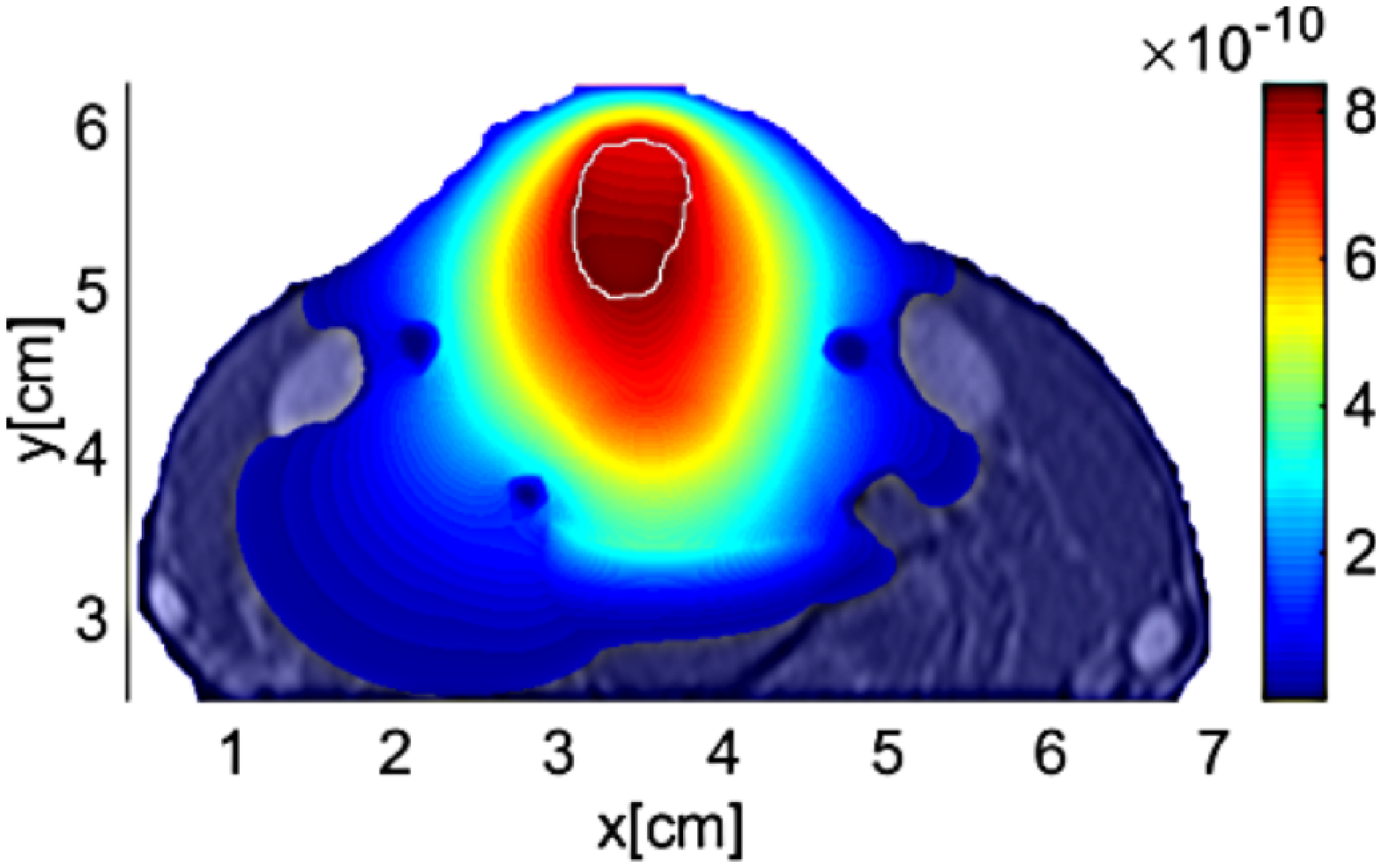}
}
\caption{Spatial distributions of $\Phi$ calculated from the RTE in the human neck model (Fig. \ref{fig1} (b)) at fixed times. Unit of the color bar is [W/cm]. The light source is represented by a red arrow in (a), and the boundary of the trachea is denoted by a white solid curve in each figure. }
\label{fig4}
\end{figure}

The developed numerical scheme of the 3UW+4RK is used to describe photon migration in the human neck model shown in Fig. \ref{fig1} (b). 
Figure \ref{fig4} shows the snapshots of $\Phi$-distributions at fixed times in the human neck model,  where light is incident on the front surface of the human neck at the position of ($x$, $y$) = (3.5 cm, 6.3 cm). 

As listed in Table \ref{tab:1}, the trachea is a void region indicated by a white solid curve in each figure of Fig. \ref{fig4}. 
Therefore, at the early time of 46.2 ps shown in Fig. \ref{fig4} (a), 
photons propagate without being scattered or absorbed inside the trachea. 
At the time of 79.9 ps shown in Fig. \ref{fig4} (b), the photons reach the trachea boundary  far from the source position, 
and migrate diffusively in the background tissue where they are scattered in all directions, 
and some portion of the scattered photons return back to the trachea as the diffusive reflection. 
Then, the reflected photons travel back to the trachea boundary near the source position as shown in Fig. \ref{fig4} (c), 
repeat diffusive reflections back and forth inside the trachea, and 
gradually migrate into the tissue outside the trachea as diffusive propagation as shown in Fig. \ref{fig4} (d) to (f). 
It looks as if the void region of the trachea acts as a light source. 

Note that the magnitudes of $\Phi$ indicated by the color bars rapidly decay as time goes and photons migrate into the tissue. 
Also note that photons are absorbed especially by veins and arteries having large $\mu_a$ which appear as dark blue spots in all of Fig. \ref{fig4} (a) to (f). 
Thus, few photons reach the backside of the human neck. 
In fact, it is confirmed that at the backside position, 
($x$, $y$) = (3.5 cm, 0.0 cm), 
the magnitude of $\Phi$ is kept at less than $10^{-10}$ during the time period from 0 ps to 800 ps. 

These results suggest that near the light source, the accurate scheme for solving the RTE is necessary, while far from the light source, the approximate and efficient scheme is good enough to describe photon migration in the human neck model. 

\section{Conclusions}

To describe photon migration in the human neck, 
we have developed the accurate and efficient numerical scheme to solve the RTE based on the 3rd order upwind and 4th order Runge-Kutta methods for the purpose of comprehensive diagnose of thyroid cancer by optical imaging. 
We have confirmed the validity of the developed scheme by comparison with the analytical solutions in  homogeneous media. 
Numerical simulations of photon migration in the human neck model have shown a complicated pattern of photon migration, 
arising from multiple diffusive reflection near the trachea. 

\section*{Acknowledgements}

This work is supported in part by Grants-in-Aid for Regional R\&D Proposal-Based
Program from Northern Advancement Center for Science \& Technology of
Hokkaido Japan, and the Japan Science and Technology Agency.  
We would like to express appreciation to Dr. Kazumichi Kobayashi for fruitful discussions. 















\end{document}